\documentclass[12pt,preprint]{aastex}
\usepackage{graphicx}
\usepackage{natbib}
\shorttitle{Adapting and expanding interferometric arrays}
\shortauthors{Karastergiou et al.}

\begin{document}
   \title{Adapting and expanding interferometric arrays}
   \author{A. Karastergiou and R. Neri}
     \affil{IRAM, 300 rue de la Piscine, Domaine Universitaire, Saint
              Martin d'H\`eres, France}
     \and 
     \author{M. A. Gurwell}
     \affil{Harvard-Smithsonian Center for Astrophysics, 60 Garden
     Street, Cambridge, MA 02138 }

\begin{abstract}
We outline here a simple yet efficient method for finding optimized
  configurations of the elements of radio-astronomical interferometers
  with fixed pad locations. The method can be successfully applied, as
  we demonstrate, to define new configurations when changes to the
  array take place. This may include the addition of new pads or new
  antennas, or the loss of pads or antennas. Our method is based on
  identifying which placement of elements provides the most
  appropriate $uv$ plane sampling for astronomical imaging.
\end{abstract}


\section{Introduction}
The imaging qualities of an interferometric telescope are determined
by the characteristics of the synthesized beam, which depend mostly on
the location of the elements of the telescope and the coordinates of
the astronomical source during an observation. These two parameters
define the positions of a set of observed Fourier samples
(visibilities) on the $uv$ plane. Various considerations in
positioning the antennas of an array are presented in a number of
articles
\citep[e.g.][]{cor86,cor88,tms91,chu93,kog97,woo99,con00a,con00b}. In
this paper, we mainly drew inspiration from the works of
\citet{Keto97} and \citet{boo01,boo02}. An important aspect of the
discussion revolves around the most appropriate distribution of the
visibilities within the $uv$ plane. The benefits of uniform $uv$
sampling for sparse interferometers are clearly pointed out in Keto
(1997). On the other hand, Boone (2002) tackles the problem in a dense
interferometer, where visibilities with a Gaussian radial distribution
achieve a more desirable Gaussian beam. In general, a refinement of
the methods of designing an interferometric radio-telescope involves
weighting the $uv$ samples according to their contribution to the
overall noise.

The problem addressed in this paper is different from the general
problem of the design of an interferometric array, although the
solution is based on the same principles. Let us assume that the
interferometer in question consists of $N_A$ antennas. The antennas
are firmly positioned on $N_A$ of $N_p$ fixed pads during an
observation, and each set of $N_A$ predefined antenna positions used
for observations is referred to hereafter as a configuration. For
instance, there may be a compact configuration, which results in a
large synthesized beam, an extended configuration with a narrow beam
and a number of intermediate configurations that can be combined to
fulfill the imaging requirements. In such an interferometer, $N_p$ is
greater than $N_A$. The problem we address is how to find an
appropriate set of configurations, given the positions of the
pads. This question may appear slightly paradoxical at first, because
it is reasonable to assume that the pads were built in an optimal way
from the start, such that the original configurations remain
appropriate. However, as we discuss throughout this paper, there are a
number of reasons for which we may like to adapt our configurations to
a change at a later stage.

Our solution to the problem stated above is based on the principle
described in the first paragraph, that the properties of the
synthesized beam of an interferometer generally depend on the
positions of the elements of the array. In Section 2 we describe the
problem and detail the steps of our method.  As applications, we
present two examples in Section 3. The first demonstrates the case
where new pads are available, as it happened at the IRAM Plateau de
Bure Interferometer \citep{gui92}. The second is related to the
Submillimeter Array \citep[SMA, ][]{hml04}, to which we have applied our
method in finding good configurations with a variable number of
antennas. A summary and concluding remarks are given in Section 4.

\section{The method}
\subsection{The principle}
The imaging quality of an interferometric array during a particular
observation depends on the array configuration and the position of the
source in the sky. This implies that the optimal configuration is
different for different sources. A grouping of the target sources by
declination may lead to the consideration of a number of
configurations optimized for each group. However, the smallest number
of total configurations of the array is generally desired, since
configuration changes are difficult and expensive tasks. 

The principle in the method of Boone for positioning antennas on a
real landscape with no other constraints, is to first choose a desired
distribution of $uv$ samples, and use this to generate pressure forces
which move antennas in a quasi-continuous way on the terrain, such
that the desired $uv$ distribution is approximated to within some
small limits. The movements are restricted only by real-life
obstacles. In our method, freedom is restricted, since antennas can
only be placed on the pads. For this reason, a direct application of
the Boone method is not possible, even though we adopt the strategy of
optimizing the configuration in the Fourier domain. Our approach is to
examine all the possible configurations, of which there is a large but
finite number, and see which one results in the most desirable
$uv$ distribution.
\subsection{The $uv$ density problem}
In order to rate each $uv$ distribution, one must first consider the
density of $uv$ samples in the plane. Boone (2002) presents detailed
arguments about the effects of the density of $uv$ samples on the
optimal $uv$ distribution. Generally, the desired $uv$ sample
distribution is Gaussian in the radial direction and uniform in the
azimuthal direction; this obviously results in a Gaussian shaped
beam. However, in the general case where there is no a priori
knowledge on the structure of the astronomical target, the conclusion
of Boone (2002) is that when the density of $uv$ samples is low, the
standard deviation of the desired radial $uv$ distribution should be
large, to guarantee that all regions of the $uv$ plane are
sampled. This suggests, that in low density interferometers where the
length of the baselines is large and the number of antennas small,
sampling as many regions of the $uv$ plane as possible takes
precedence over Gaussian radial sampling. Typical baseline sizes for
obtaining Gaussian $uv$ coverage at various interferometric telescopes
are given in Table 1 of Boone (2002), from which it is apparent that
arrays such as the IRAM interferometer and the SMA operate in a low
$uv$ density regime.

If the $uv$ sampling aimed for is radially Gaussian, then the method
suggested by Boone (2001), to evaluate how ``Gaussian'' the
$uv$ sample distribution is, applies well. In short, one divides the
$uv$ plane into regions by a number of equally spaced azimuthal cuts
and a number of circles at radii such that there is equal probability
of having a constant number of samples in each sector if the samples
are 2-d Gaussian distributed. A complete presentation of this
evaluation method can be found in Boone (2001).

\subsection{The low density solution}

In the case where the $uv$ samples are sparsely distributed in the
Fourier plane, the above method cannot differentiate adequately
between two configurations that lead to substantially different
$uv$ distributions. Dividing the $uv$ plane into more sectors does not
solve the problem, but only results in many sectors void of $uv$
samples. 

We have devised a solution to this problem that characterizes each
configuration by a single quantity $F$, related to the quality of the
resulting $uv$ distribution and according to which a ranking is
possible. During a typical observation using Earth-rotation synthesis,
an array of $N_A$ elements observes a number of visibilities $N_V$,
given by $N_A*(N_A-1)$ multiplied by the number of time integration
intervals.  Then $F$ is given by the simple equation:
\begin{equation}
F=R^2\sum_{i=1}^{N_V-1}\sum_{j=i+1}^{N_V}\frac{1}{|\mbox{\boldmath$
    V_i$} - \mbox{\boldmath$V_j$}|^2}.
\end{equation}
{\boldmath $V_i$} denotes the position $[u,v]$ of a particular sample,
and $R$ the size of the longest baseline. The rationale of this
equation is based on the simple analogy between the $uv$ plane and a
closed, two dimensional surface, in which the samples resemble charged
particles. The charged particles under such conditions would attempt
to distribute themselves such that the energy of the system would be
minimal, achieving more or less equal density all over the surface.
According to this analogy, we search for the distribution that results
in the lowest value of $F$. The normalisation factor $R^2$ is applied
to permit the comparison between more compact and more extended
configurations. The ``forces'' in the above analogy are applied on the
$uv$ samples as opposed to Boone's ``forces'' which are applied
directly to the antenna stations. \citet{cor88} proposed a similar
method to design interferometric arrays from scratch, which is an
altogether different problem to the one we address here, as we have
already stressed in the introduction. A brief discussion on that
method can be found in \citet{Keto97}. There, its main weakness is
pointed out, i.e. that the ``forces'' tend to place antennas on the
boundary of their domain, which does not result in an optimal uniform
$uv$ coverage. As we demonstrate in this work, our own method is used
to identify the most appropriate set of fixed pads of an already
existing interferometer, which eliminates the aforementioned
possibility of antenna placement.

The code we have developed to identify the best configurations
according to the above criterion in the low density case, goes through
the following steps:
\begin{enumerate}
\item Three input files are read into the program. The first contains
  the geocentric coordinates of all the available pads, in an $xyz$
  list. The second input file contains one line for each antenna we
  would like to place on the array. Each line consists of pairs of
  indices, defining ranges of pads where the antenna may be
  positioned. The indices originate from the first input file. The
  third file contains the parameters of the planned observations, such
  as the declination of the source and the duration of the
  observations for synthesis imaging.
\item Having read in the input files, the number of antennas is now
  known. Each antenna is placed on the first pad of the allowed ranges
  and the baselines are calculated. For a given pair of antennas, two
  baselines are computed, as the difference in antenna position in
  both directions.
\item According to the observation parameters, the $uv$ samples due to
  Earth rotation are calculated for the source of a particular
  declination.
\item The parameter $F$ is computed for the sum of the $uv$ samples,
  which are $N_V$ (as defined above), in total.
\item The first antenna of the list is moved to the next pad, within
  the desired range. The new baselines are computed and rotated due to
  the Earth's motion. The new $uv$ samples then replace those relating
  to the antenna which was moved, and step 4 is repeated. If the
  parameter $F$ has decreased compared to its previous value, the
  current configuration is stored as the temporary solution.
\item When the first antenna is moved beyond the last possible
  position in its range, it is moved back to the first position and at
  the same time the second antenna is moved to the second
  position. While the number of computations is larger when more than
  one antenna is moved to a different pad at once, this occurs only a
  small fraction of the time. Mostly the changes involve one antenna
  and therefore only $N_A-1$ baselines.
\item When all antenna ranges have been considered, the configuration
  stored is the one with the lowest $F$ value. Only the current
  solution is kept in the computer memory every time, and not all the
  computed $F$ values. This generally precludes the possibility to
  compare the solution with the neighbouring values of $F$, as well as
  with other local minima. However, to include this in our code
  requires an unfeasible amount of computer memory, or writing out the
  solutions to the hard-disk, which slows up the procedure very much.

\end{enumerate}

In applying the above procedure, one or more of the antennas may be
fixed to particular positions. Such constraints may stem from the
desire to include particular stations or baselines to the array, for
practical purposes or purposes of calibration. In the case where a
number of configurations are optimized, as discussed in the following
section, a smooth change in the synthesized beam size may also lead to
constraints in the antenna positioning. Also, the function given above
implies equal weighting for all visibilities. We have experimented
with weighting that accounts in a crude manner for the atmospheric
thermal noise at different elevations, by simulating standard and
non-standard observations. Higher weights were given to the Fourier
samples collected around the transit of the simulated source, when the
atmospheric thermal noise is normally lowest. We have found no
examples were this process alters the preferred solution.

In the high $uv$ density case, the same process can be applied, only
the evaluation criterion will be different, as already mentioned.

The number of iterations $N_i$ depends on the number of
antennas $N_A$ which move, the number of pads that can host these
antennas $N_P$ and the number of visibilities in the simulated
observation $N_V$ as
\begin{equation}
N_i\approx N_V^2*\frac{N_P!}{(N_P-N_A)!}
\end{equation}
This equation is approximative and assumes that each of the antennas
that are moved has an equal number of possible host stations, which is
not necessarily always true. The large number of possibilities the
software has to loop through suggests a large execution time. However,
having optimized the code to perform the bare minimum calculations per
step of the loop to increase overall performance, we have obtained
results for the problems presented here within minutes. Considering
the fact that an optimization of the configurations of an
interferometric array is not a task that will be necessarily repeated
very often, the time requirements of our method are well within
acceptable limits.

\subsection{Optimizing multiple configurations}

In low density interferometers a combination of configurations of
different size is often used to better sample the $uv$ plane. The
observations are carried out in two or more configurations in
succession and the resulting visibilities are combined in the imaging
process. In such cases, the imaging qualities of the interferometer
must be good for the combinations of configurations used. The
possibility for such optimization can be easily added to the procedure
described above: to optimizing a pair of configurations to match each
other, one configuration is optimized alone first, and then the
resulting visibilities are injected in the $uv$ plane for the
optimization of the second configuration. In calculating $F$, there
are then twice the number of visibilities, half of which remain
constant throughout the process as they refer to the already optimized
configuration. The second configuration is then optimized to
complement the first. Constraints on the second configuration can be
placed in terms of the available stations for positioning the
antennas.

\section{Example applications}
\subsection{Optimizing new configurations for the IRAM interferometer}
\begin{figure}
\centerline{ \includegraphics[width=11cm, angle=-90]{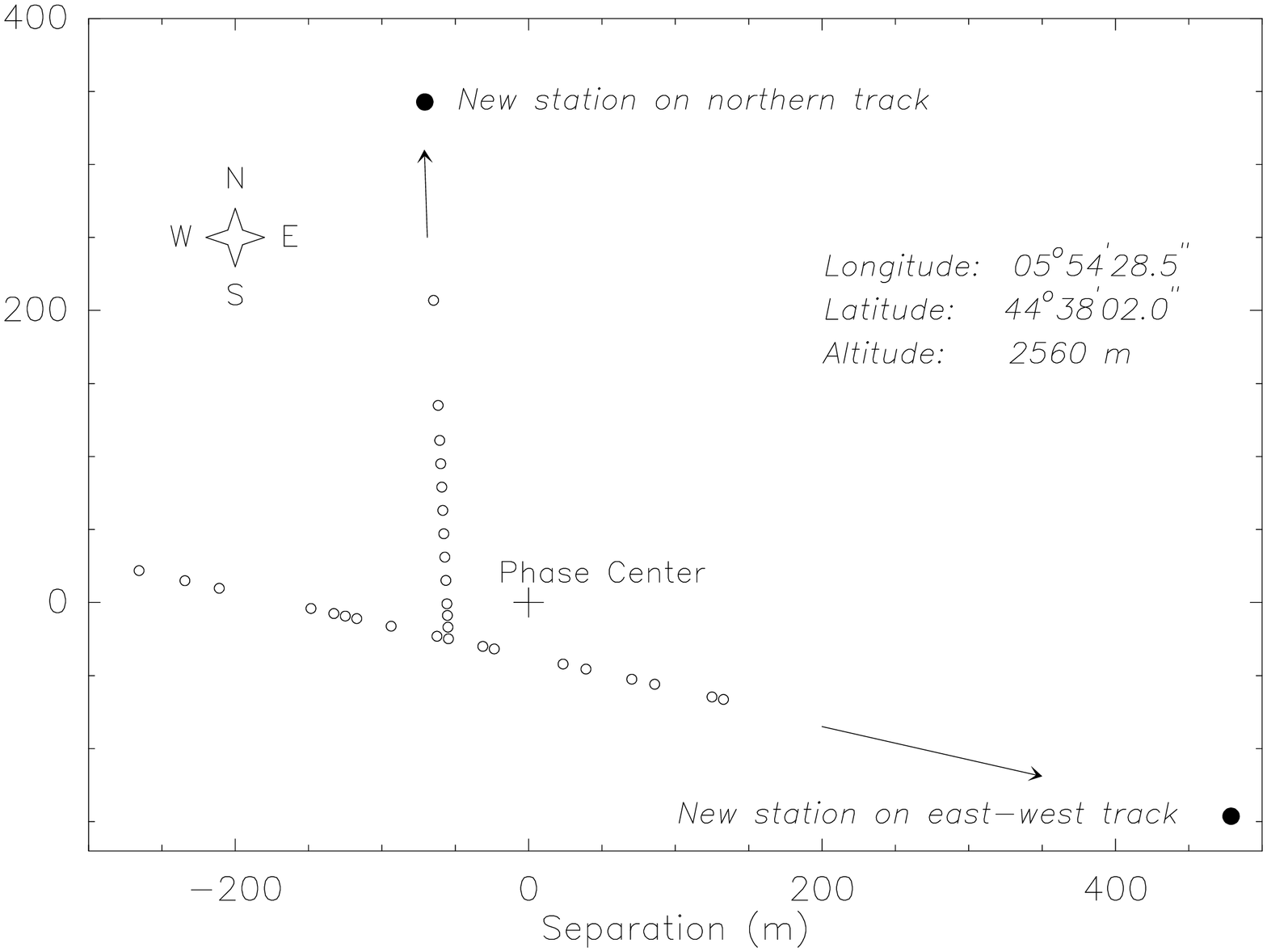}}
  \caption{The station positions of the IRAM interferometer. The two
  new stations at the far north and far east of the tracks are
  emphasized, and the arrows denote the track extensions. The latitude
  and longitude refer to the phase center of the array.}
  \label{pdbi}
\end{figure}
The IRAM\footnote{IRAM is supported by the CNRS (Centre National de la
Recherche Scientifique, France), the MPG (Max Planck Gesellschaft,
Germany), and the IGN (Instituto Geogr\`afico Nacional, Spain)}
interferometer is located in the South of the French Alps, at an
altitude of $\approx 2550$ m. Currently the IRAM interferometer
comprises $N_A=6$ identical antennas of 15-m diameter, that can be
positioned on $N_p=30$ pads along a {\it T}-shaped track. Until
recently, the north-south arm was 232 m long, and the almost east-west
oriented arm extended 216~m west and 192 m east of the
intersection. The angle between the arms is $75\degr$. In order to
better sample the $uv$ plane, the IRAM interferometer operates in four
different configurations, ranging between compact and extended
baselines. The choice of the total number of configurations and the
design of each of them has been heavily influenced in the past, by the
very demanding requirements of antenna movement in winter
conditions. This currently takes place two to three times in each of
the 6-month observing terms.

At the end of the summer of 2005, the northern track was extended to
368~m and one new pad has been constructed at the end of this
track. Also, the east-west track was extended from 408~m to 760~m,
with one new pad at the far eastern end of the extension, bringing the
total number of pads to $N_p=32$. A diagram showing the positions of
the current and new stations is shown in Figure \ref{pdbi}. The new
stations can be clearly identified at the easternmost and northernmost
positions of the tracks.  Despite the track extensions, the total
number of configurations will remain constant, due to the practical
difficulties of moving antennas on a plateau at the top of a snowy
mountain. However, the 4 new configurations and combinations of
consecutive configurations have been designed with an emphasis on the
quality of the synthesized beams. The first constraint for designing
the new configurations, therefore, is to preserve the total number of
4, and optimize consecutive configurations to match each-other well.

The strategy we followed to optimize all 4 configurations consists of
these steps:
\begin{itemize}
\item Optimize the most extended configuration ({\bf A}), which
  includes the farthest stations to the north, east and west.
\item Optimize a second extended configuration ({\bf B}) which
  includes the far northern and western stations, but where the most
  eastern antenna is limited to shorter distances. This configuration
  is optimized to match the first configuration as described in the
  previous section. As a result, the combination {\bf AB} has optimal
  imaging qualities.
\item Optimize a very compact configuration ({\bf D}), taking care
  about shadowing effects and obviously avoiding antenna collisions.
\item Optimize a second compact configuration ({\bf C}) by
  constraining the available stations according to both {\bf B} and
  {\bf D}. The {\bf C} configuration is optimized in the $uv$ plane by
  combining it with {\bf B}, however the quality of the {\bf CD}
  combination is also checked.
\end{itemize}

As a demonstration we show here the solution for the most extended
configuration {\bf A}. To begin, we choose a particular declination
for which we would like to optimize the configuration. In doing so, we
consider the average declination of the sources observed at the
telescope. Also, and quite importantly, configurations that are
optimized for low declination sources (which do not rise beyond a
certain elevation) are better for sources of all declinations than
configurations which are optimized for high declinations. We have
found that optimizing for a declination of $+20\degr$ proves a good
compromise for sources of various declinations at the site. Also, the
tracking time of the Earth rotation synthesis used for the
optimization process was 9 hours.

\begin{figure}
\centerline{ \includegraphics[width=9cm, angle=-90]{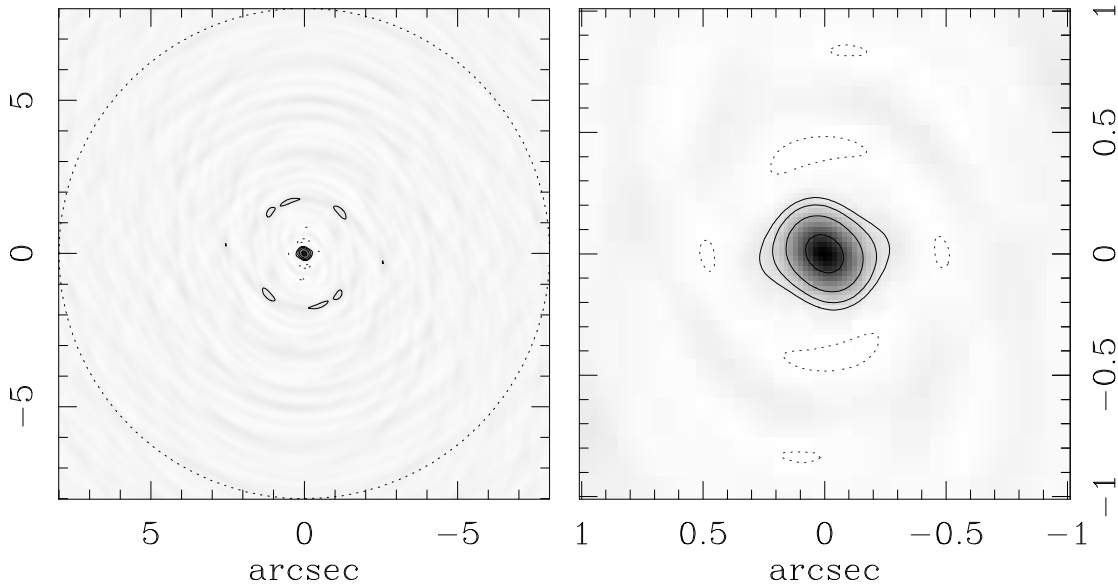}}
  \caption{ The synthesized beam obtained by a 9-hour observation of a
    source at $+45\degr$ declination, using the new stations of the
    Plateau de Bure interferometer, in the most extended configuration
    (A). The image on the left has approximately the size of the
    primary beam at 300 GHz (dashed circle), whereas on the right we
    have zoomed in on the central region. The contour levels at
    $\pm$10\%, 20\%, 40\% and 80\% measure the sidelobe levels.}

  \label{compA}
\end{figure}
\begin{figure}
\centerline{
  \includegraphics[width=14cm]{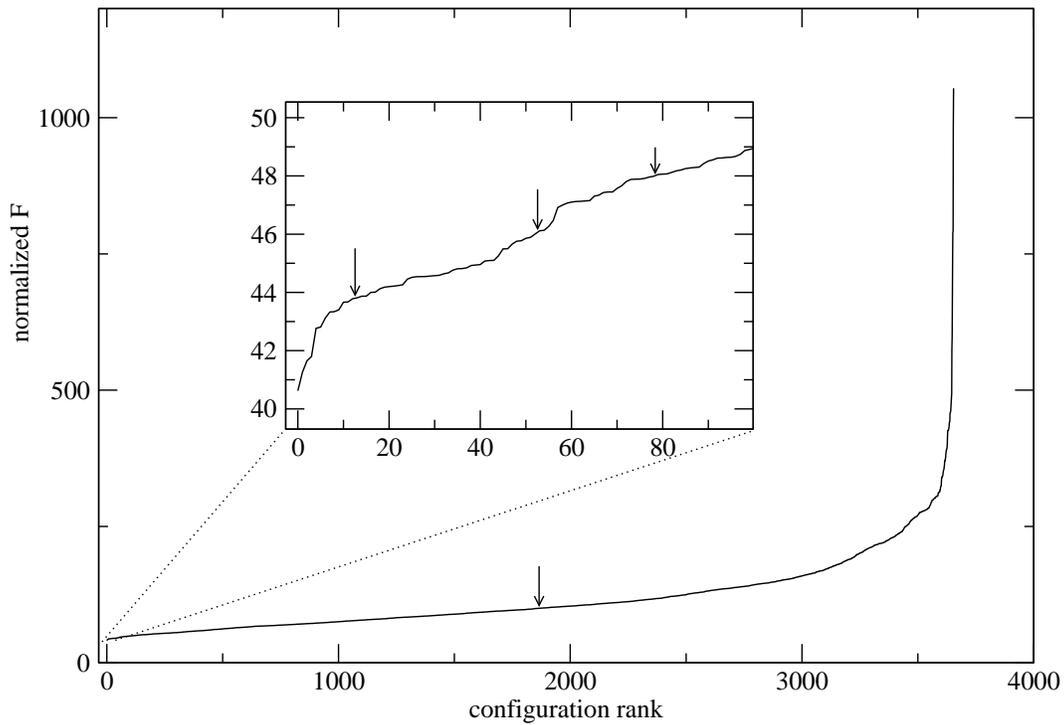}}
  \caption{For the optimization of the most extended configuration of
  the IRAM interferometer, we show the $F$ parameter versus the final
  ranking of all the possibilities the code calculated. The highest
  ranking configurations are shown in the inset, where $F$ is
  lowest. The beams and $uv$ coverage of the 4 marked points are shown
  in Fig. \ref{converge}.}

  \label{paraF}
\end{figure}
\begin{figure}
\centerline{
  \includegraphics[width=14cm]{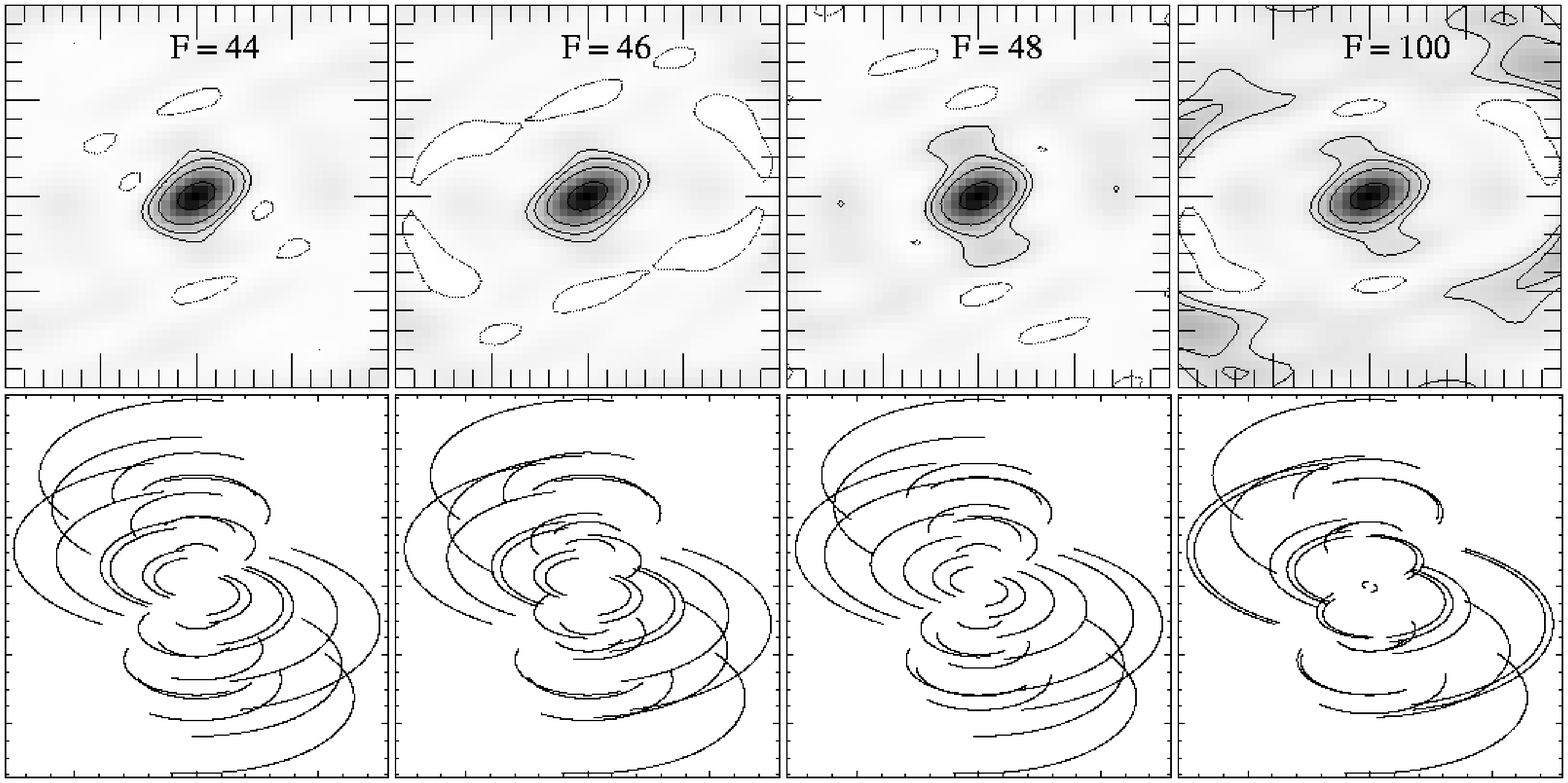}}
  \caption{A demonstration of the convergence of our method, moving
   from right to left. The beams from 4 configurations sampled from
   Fig. \ref{paraF}, for a source at declination $+20\degr$. The scale
   of the beam is that of Fig. \ref{compA}. The $uv$ plane has a
   radius of 760~m.}
  \label{converge}
\end{figure}

Figure \ref{compA} shows the beam achieved by the new optimized {\bf
A} configuration of the IRAM interferometer. The beam shown results
from a 9-hour observation of a $+45\degr$ declination source per
configuration. It demonstrates also that despite the optimization
being carried out for $+20\degr$ declination sources, the quality
remains good at higher declinations. The two panels of the figure are
shown at a scale of the primary and synthesized beams
(left/right). Our method not only results in a beam with good imaging
properties in an absolute sense, but also results in a rounder beam
with lower sidelobes compared to the previous most extended
configuration in this array.

For the specific optimization described here, there is only a limited
number of possible antenna moves, since three antennas were anchored
to the most extended positions. For that reason it was feasible to
write to disk all the values of $F$ during the search.  We have ranked
all the configurations that our method has considered in the above
optimization process, according to the $F$ parameter, with the best
configuration (minimum $F$) assigned rank 1. In Fig. \ref{paraF} we
plot $F$ versus the rank (here we have normalized $F$ by
$N_V(N_V-1)$). Although the curve is quite flat up to rank $\approx
3600$, a visual inspection of the beams shows that the quality
deteriorates gradually but noticeably. The inset of Fig. \ref{paraF}
shows the best configurations according to our criterion. The 4 marks
on this figure denote the configurations which we use to demonstrate
the convergence of our method. In Fig. \ref{converge} we show the
beams and $uv$ coverage of these four configurations. The quality of
the beam obviously improves in terms of sidelobe levels from right to
left. Although it is hard to rank these examples in terms of their
$uv$ coverage by a simple visual inspection, our method identifies the
left-most case as most ``uniform''. The left most example also
produces the lowest sidelobe levels. This is a clear demonstration
that the optimization in the $uv$ plane results in a good quality beam
in the image plane.

\subsection{Adapting the very extended array of the SMA}
\begin{figure}
\centerline{ \includegraphics[width=13cm, angle=-90]{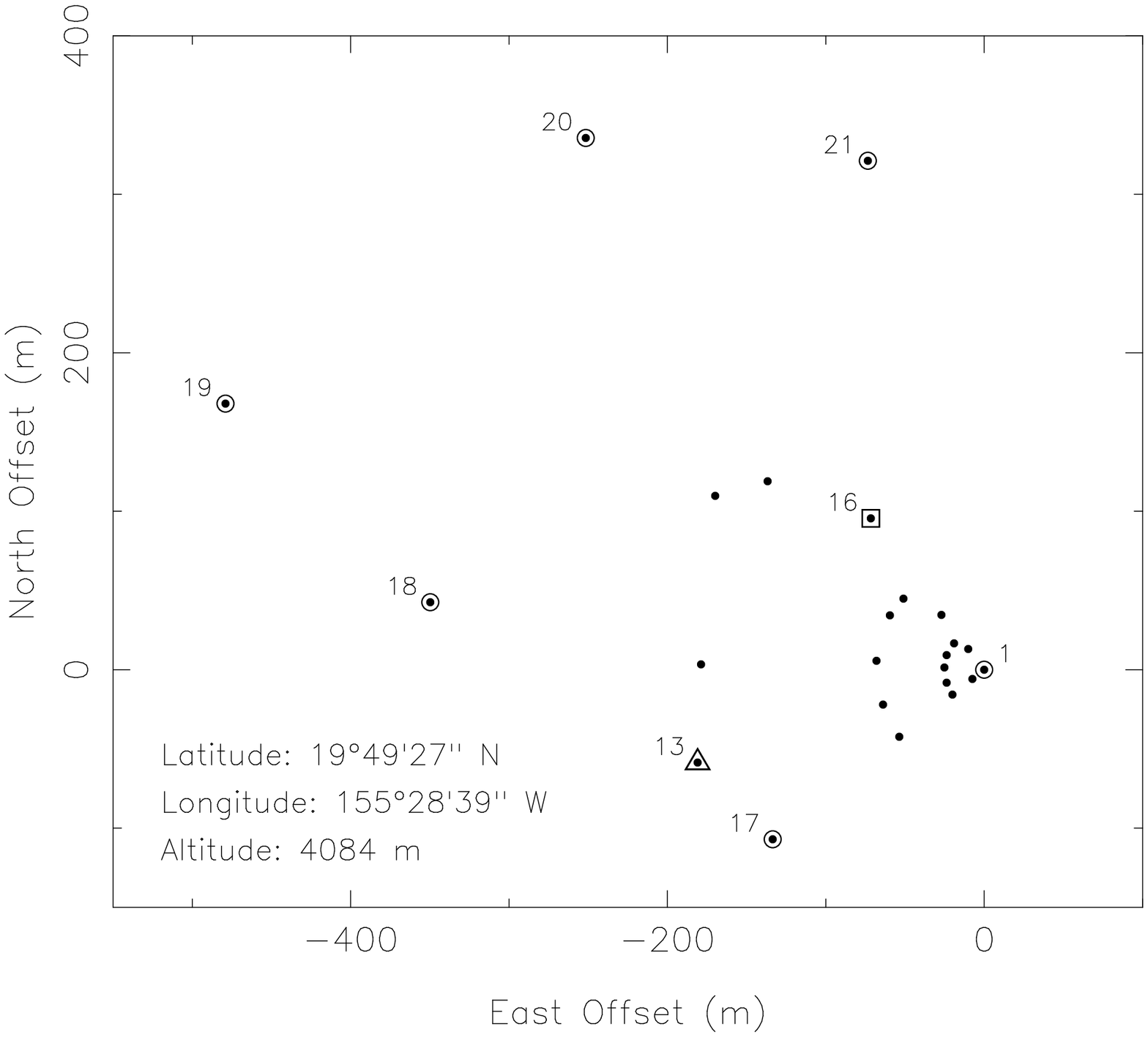}}
  \caption{The location of the 24 pads of the SMA. The circles denote
  the chosen pads for a six element, very extended array. The square
  and triangle show the pads identified in this work, which optimize
  this array in the case of seven and eight elements respectively. The
  phase center is at pad 1.}

  \label{SMA}
\end{figure}

The Submillimeter Array (SMA)\footnote{The SMA is a collaborative
project of the Smithsonian Astrophysical Observatory of the
Smithsonian Institution and the Academia Sinica Institute of Astronomy
and Astrophysics of Taiwan.} is located near the summit of Mauna Kea,
on the Island of Hawaii, at an altitude near 4080 m.  The SMA is
comprised of eight 6-m diameter submillimeter-capable radio
telescopes.  The initial Smithsonian concept for the SMA was for six
telescopes, but in 1996 the Academia Sinica Institute of Astronomy and
Astrophysics (ASIAA) joined the project by agreeing to add two more
telescopes \citep{hml04}.

There are currently 24 pad locations for the SMA, with the first 21
determined when the array was scheduled to consist of six elements.
The locations were based on the work of Keto (1997), including the use
of the Reuleaux triangles for optimization.  The scheme utilizes four
nested "rings" of pads (see Fig. \ref{SMA}), with each ring providing a
different spatial coverage and resolution.  The addition of the two
ASIAA telescopes was accomodated within the Reuleaux triangle scheme
on an ad hoc basis, resulting in an additional three pads for
configurations within the three smallest nested rings (Ho, Moran, and
Lo 2004).

The highest spatial resolution array (called the 'very extended array'
and denoted in Fig. \ref{SMA} as SMA pads 1, 17, 18, 19, 20, and 21)
did not have additional pads added, and relies on the use of
preexisting pads.  This situation provides an opportunity to apply the
concepts developed in this work.  The specific question to be
addressed is: given the original six pad locations for the very
extended array, which two additional pads optimize the $uv$ coverage
while minimizing degradation to spatial resolution.  A secondary,
related question is which pads are best if there are less than the
full complement of eight telescopes (for example, if one is undergoing
renovation or servicing).
\begin{figure}
\centerline{
  \includegraphics[width=14cm, angle=-90]{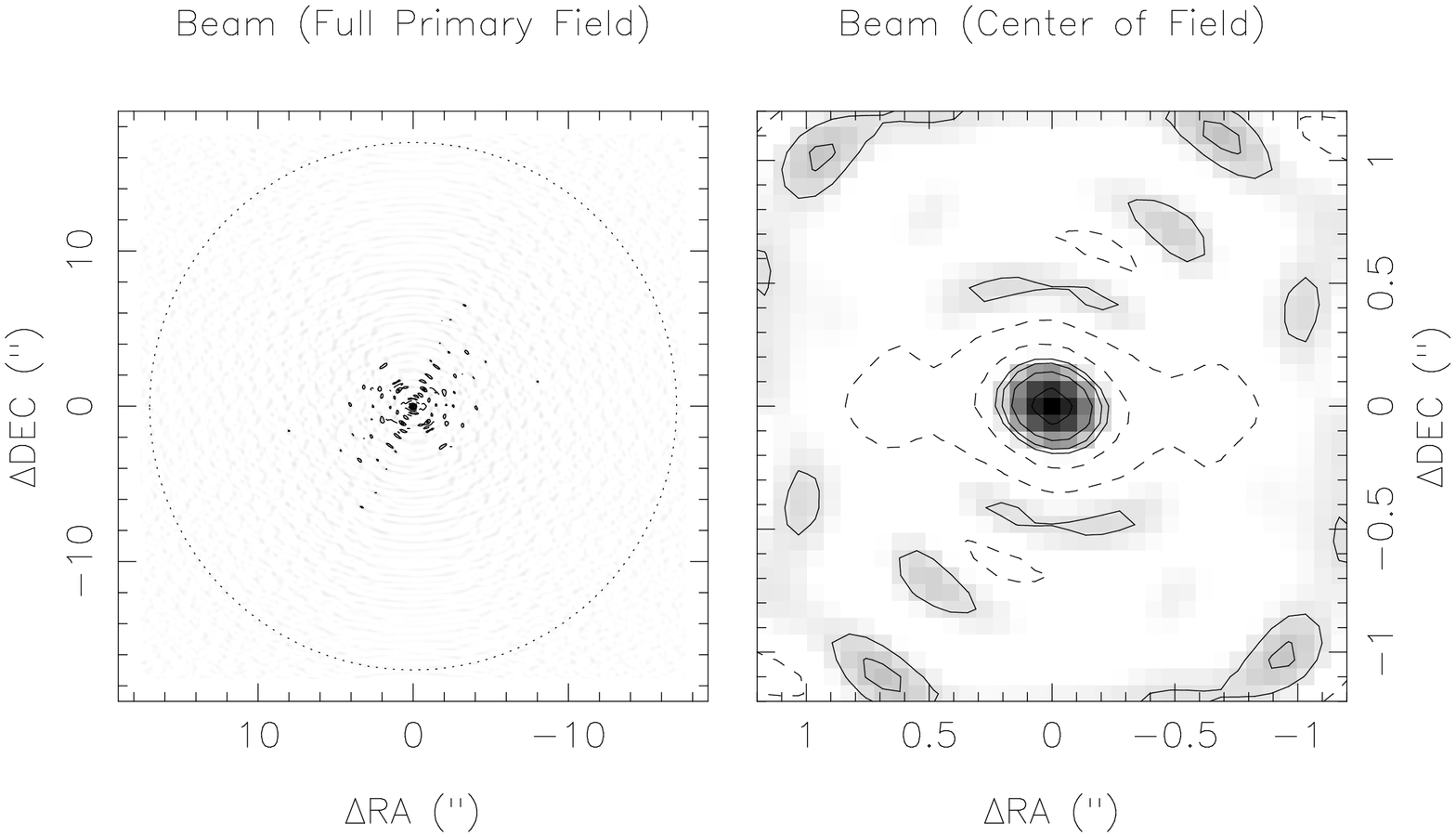}}
  \caption{The synthesized beam of the six element, very extended SMA
  array, for a source at a declination of $+30\degr$, shown at the
  level of the primary field and the size of the synthesized
  beam. Countour lines are at the same levels as in Fig. 2. }
  \label{SMAbeams1}
\end{figure}
\begin{figure}
\centerline{
  \includegraphics[width=14cm, angle=-90]{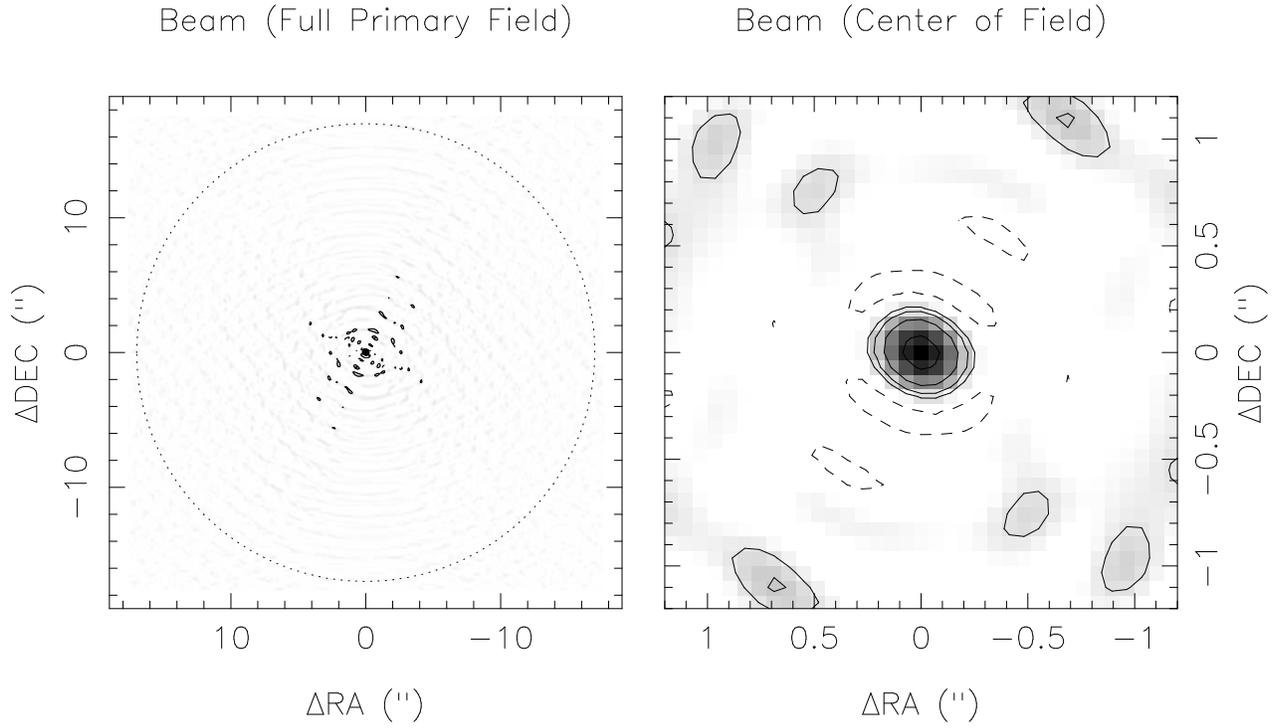}}
  \caption{The synthesized beam of the very extended SMA array, with a
  total of 7 antennas. The beam retains its shape and the sidelobes
  are supressed compared to the six element array.}
  \label{SMAbeams2}
\end{figure}
\begin{figure}
\centerline{
  \includegraphics[width=14cm, angle=-90]{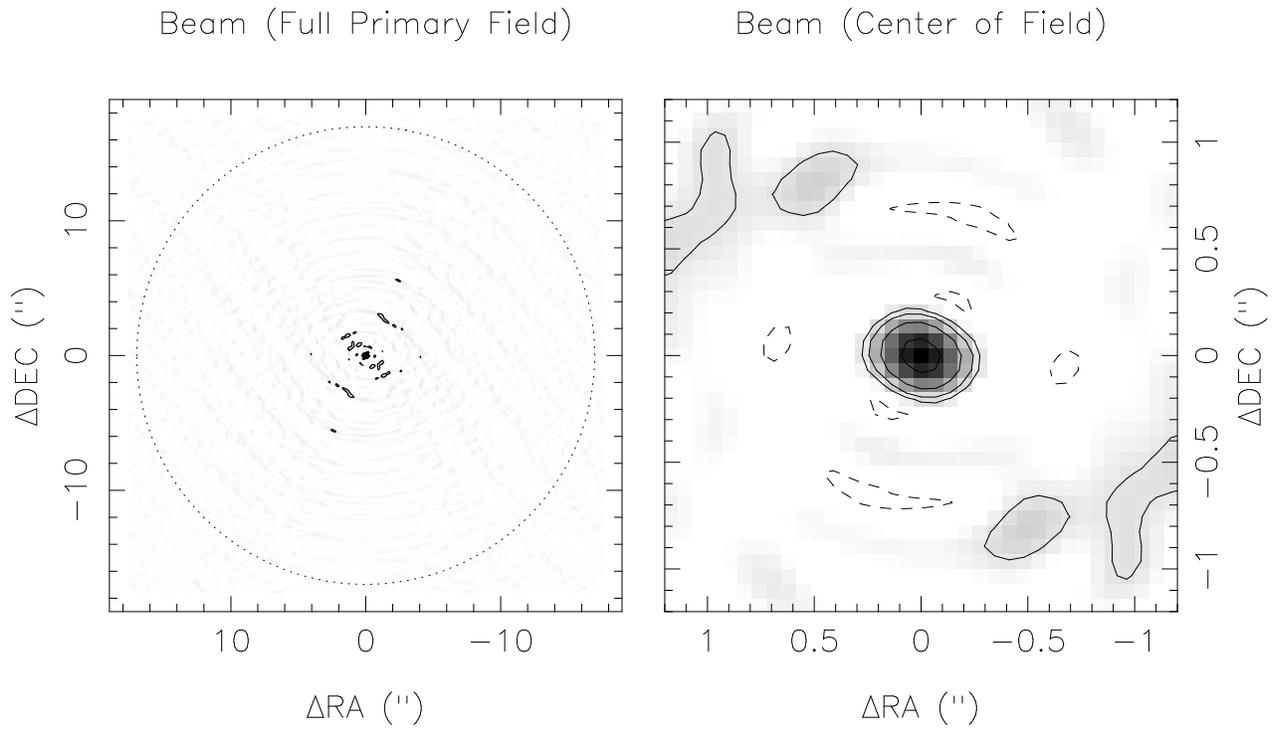}}
  \caption{The synthesized beam of the very extended SMA array, with a
  total of 8 antennas. The sidelobes are further supressed compared to
  the six and seven element array.}
  \label{SMAbeams3}
\end{figure}

The methods used in this work were applied to address these questions,
and we find that if seven telescopes are available, the addition of
pad 16 to the original very extended configuration provides the best
$uv$ coverage, and if eight telescopes are available, the addition of
pads 13 and 16 together is best.  This optimization appears to hold
true for a wide variety of source declinations, though does change
slightly for very low (below $-25\degr$) declinations. Figure
\ref{SMAbeams1} shows the synthesized beam obtained for the original
six element configuration, for a declination of $+30\degr$ at 345
GHz. Figures \ref{SMAbeams2} and \ref{SMAbeams3} show the improvement
in the beam (primarily in suppression of sidelobes) for the seven
element and eight element configurations, respectively. It is worth
noting that the solution for the seven element case obtained rapidly
here, was previously determined by SMA staff through a laborious
search of possibilities using their synthesis beam generation
tools. We consider the agreement in the two methods as confirmation
that both correctly address the problem at hand, despite the fact that
they are completely different in approach.

\section{Conclusions}
We have applied a simple principle, based on the ideas of Keto (1997)
and Boone (2001, 2002), to develop a procedure that identifies
configurations with good $uv$ coverage for low density interferometers
with defined pads. The method has been put to the test in the
particular situation where new antenna pads are available at the IRAM
interferometer, which require a new set of antenna configurations. It
has also been applied to identify configurations for the SMA with
different numbers of antennas. Gaining or losing pads or antennas are
events that are by no means unlikely for interferometric telescopes
that apply multiple configurations.

The total number of configurations in which any interferometer
observes is defined not only by scientific principles, but also by
practical constraints. If the total number of possible configurations
changes at some point in time, a new entire set of configurations will
need to be redefined, and the method presented here is suitable to
solve this problem in a fast and elegant way.

In summary, we propose here a fast method for optimizing the positions
of antennas in an interferometric array with defined pads, as opposed
to the method of Boone where antennas can be moved around on a real
landscape. Our method resembles the method of Boone in that it
consists of a repetition of a two step process: changes are made to
antenna positions and the new configuration is evaluated. For the
first part, our parameter space is finite so we explore it entirely,
unlike the Boone method which explores part of the infinite space and
converges to one of many possible solutions. Regarding the second step
of the process, our method can be applied to interferometers with a
low density of $uv$ samples, in contrast to the method of Boone, which
only works for high density. Finally, a two-step procedure to define
configurations in high density interferometers with defined pads can
be developped, where the first step is taken from the method described
here, and the second from the method of Boone. That would involve
looping through all the possible configurations and ranking them by
means of the Gaussian method described in Section 2.2. 
\bibliography{journals,iram}
\end{document}